\magnification=1200
\baselineskip=18truept
\input epsf

\def\preprint{Y}
\def\draftversion{N}
\def\cap{\hsize=4.6in}

\if \draftversion Y


\fi
\def\figure#1#2#3{\vskip .2in
\if \preprint Y \midinsert \epsfxsize=#3truein
\centerline{\epsffile{figure_#1_eps}} \halign{##\hfill\quad
&\vtop{\parindent=0pt \hsize=5.5in \strut## \strut}\cr {\bf Figure
#1}&#2 \cr} \endinsert \fi}

\def\figureb#1#2{\if \preprint N \midinsert \epsfxsize=#2truein
\centerline{\epsffile{figure_#1_eps}} \halign{##\hfill\quad
&\vtop{\parindent=0pt \hsize=5.5in \strut## \strut}\cr \cr \cr
\cr \cr \cr  {\bf Figure #1} \cr} \endinsert \fi}

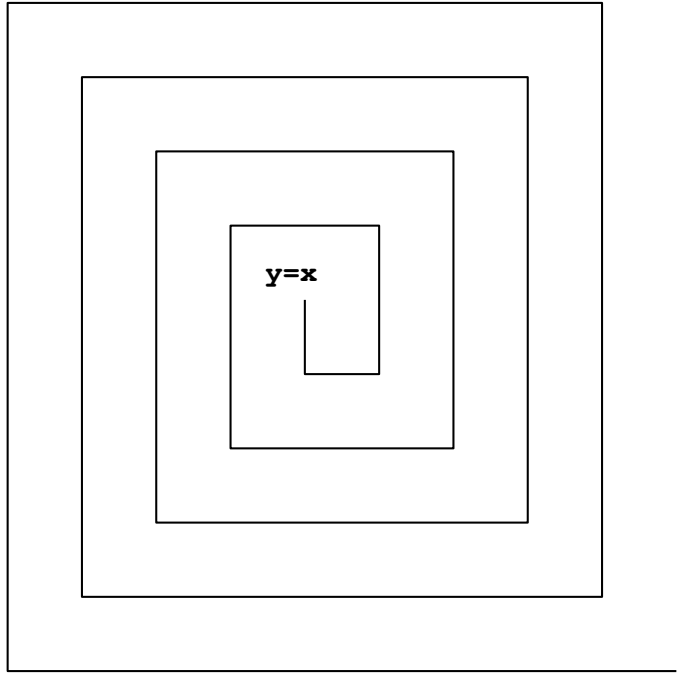
\def\captionone{\cap Axial gauge tree for constructing $\chi_\rho (x,y;A)$.}

\def\Journal#1#2#3#4{{#1} {\bf #2}, #3 (#4)}


\def\NPB{{\it  Nucl. Phys.} B}
\def\PLB{{\it  Phys. Lett.}  B}
\def\PRL{\it  Phys. Rev. Lett.}
\def\PRD{{\it  Phys. Rev.} D}

\def\ANPH{{\it Annals of Physics}}
\def\NPROC{\it Nucl. Phys. Proc. Suppl.} 
\def\JPSJ{\it J. Phys. Soc. Jpn.}
\def\ovlapa{1}
\def\luscher{2}
\def\ovlapb{3}
\def\dubna{4}
\def\kaplan{5}
\def\fands{6}
\def\candh{7}
\def\thooft{8}
\def\zinn{9}
\def\preskill{10}
\def\boulder{11}
\def\bsimon{12}
\def\geom{13}
\def\berry{14}
\def\mybounds{15}
\def\gw{16}
\def\daemi{17}
\def\kato{18}
\def\chiuone{19}
\def\fuji{20}
\def\shamir{21}
\def\ovlapdirac{22}
\def\fujiwara{23}
\def\fnn{24}
\def\higgsbound{25}
\def\latticebrs{26}

\line{\hfill RUHN-99--8}
\vskip .5cm
\centerline {\bf Noncompact chiral $U(1)$ gauge theories on the lattice.}
\vskip 1cm
\centerline{Herbert Neuberger}
\vskip .25cm
\centerline{\tt neuberg@physics.rutgers.edu}
\vskip 1.5cm
\centerline{\it Department of Physics and Astronomy}
\centerline{\it Rutgers University}
\centerline{\it Piscataway, NJ 08855--0849}
\vskip 2cm
\centerline{\bf Abstract}
\vskip 2cm
A new, adiabatic phase choice is adopted for the overlap in the
case of an infinite volume, noncompact abelian chiral gauge theory.
This gauge choice obeys the same symmetries as the Brillouin-Wigner
(BW) phase choice, and, in addition, produces a Wess-Zumino
functional that is linear in the gauge variables on the lattice. 
As a result, there are no gauge violations on the trivial
orbit in all theories, consistent and covariant anomalies are simply
related and Berry's curvature now appears as a Schwinger term.
The adiabatic phase choice can be further improved to produce
a perfect phase choice, with a lattice Wess-Zumino functional that is
just as simple as the one in continuum. When perturbative anomalies cancel,
gauge invariance in the fermionic sector is fully restored. 
The lattice 
effective action describing an anomalous abelian gauge theory
has an explicit form, close to one analyzed in the past in a 
perturbative continuum framework. 

\vfill\eject
\leftline{{\bf 1. Introduction}}
\vskip .5cm
The overlap is a general procedure to regulate chiral
gauge theories which also naturally fits on the lattice [\ovlapa]. 
At the moment this procedure is unique;
the so called ``Ginsparg-Wilson''
approach\footnote{${}^{f_1}$}{
For a recent review see [\luscher].} 
is essentially identical [\ovlapb,\dubna]. The overlap is based on the
domain wall approach of Kaplan [\kaplan] and on the infinite fermion approach
of Frolov and Slavnov [\fands]. The infinite number of fermions
reside along the extra dimension of the $d+1$ dimensional
space into which the $d$ dimensional domain
wall is embedded. For this setup to fit into the Frolov Slavnov 
approach one needs to make the gauge fields purely $d$ dimensional,
merely Xeroxed into the extra dimension [\ovlapa]. The domain wall set up of
Kaplan inherits an appealing physical picture from the continuum
domain wall setup of Callan and Harvey [\candh]. This physical picture
will be exploited later on. 

In this paper I focus on abelian chiral gauge theories. As long
as one does not think about embedding the abelian group into a
nonabelian one, one is free to consider a theory with noncompact
gauge group and unquantized charges. That this might be useful on
the lattice has been emphasized by 't Hooft [\thooft]. To avoid any other
issues related to charge quantization, this work addresses
an infinite lattice, rather than a finite one. It is simply
assumed that a reasonable thermodynamic limit exists; I made no attempts
to control it rigorously. 

The vectorial version of the class of models we consider is analyzed
in almost any field theory textbook. Here I shall use a gauged fixed
version with finite photon mass (see equation (2.5) for a lattice
transcription), and assume familiarity with some
basic facts, for example as explained in J. Zinn-Justin's textbook [\zinn]
on Euclidean Field Theory.  Mainly, 
I am relying on the fact that, as a result of the gauge breaking
terms being purely quadratic, gauge breaking effects can be 
controlled throughout the process of renormalization,
and gauge invariance continues to play a crucial role in determining
(within perturbation theory) the physical content of the continuum
theory. Therefore, 
although we shall be working in a fixed
gauge as above, but now in the chiral case, the issue
of anomalies and their cancelation still plays a central role. 

Since the models strictly speaking do not exist as continuum nontrivial
field theories, the appropriate framework is that of effective Lagrangians.
The material I need is presented in the abelian section of a paper by
J. Preskill [\preskill]. (The abelian section of that paper is joint work of
J. Preskill and M. Wise.) An anomalous chiral abelian gauge theory
differs from one in which anomalies cancel by the range of energies
it is applicable to, and by how small the photon mass can be made in
a natural way. When anomalies cancel it is natural to make the
photon massless. 

A technical simplification I shall employ has been mentioned
already in a talk I gave at Lattice'98 [\boulder]. It amounts to
replacing throughout the construction the
Wilson Dirac Hamiltonian $H_W (A)$ by its sign function.
The overlap is invariant under any replacement of $H_W (A)$
by a monotonic function of $H_W (A)$, $f(H_W (A))$. In addition
to being monotonic $f$ must be smooth and vanish at zero. 
Although the sign function is not smooth at zero, if the spectrum
of $H_W (A)$ is excluded from a vicinity of zero, one is allowed
to take $f$ as the sign function. This simplification makes
some formulae look simpler, because all energy denominators
are now simple integers, on account of the sign function
attaining only the values $\pm 1$. This simplification is merely
technical. 

The plan of the paper follows: After setting up notations and conventions
(section 2)
the main new ingredient of this paper is presented. It consists of 
a new phase choice for the overlap, an adiabatic phase choice [\bsimon]
(section 3).
The main advantage of the new phase choice is that the Wess-Zumino functional,
which measures the residual gauge dependence in the fermion determinant, 
is linear in the gauge variables (section 4). The linearity is an exact property
holding at finite lattice spacing. The previously adopted BW phase choice
yielded a Wess-Zumino functional that was nonlinear in the gauge
variables. The old and new phase choices are related 
by a phase redefinition, explained
later in the paper (sections 12 and 13). 
Two important consequence of the linearity
of the Wess-Zumino action are highlighted: absence of gauge dependence
on the trivial orbit (a property that {\it was} true of the BW phase choice,
but only in two dimensions [\ovlapb]) 
and a simple relation between the covariant and
the consistent anomaly functionals, on the lattice
(sections 5 and 6). From a previous
paper on the geometrical aspects of the overlap [\geom] it is known that
the {\it difference} between the consistent and covariant anomalies is
controlled in the overlap by a geometrical object, namely the
Berry curvature [\berry] over the space of gauge fields. This curvature
is associated with the fermionic ground state that plays a central role
in the overlap. Because with the new phase choice
the difference of the anomalies is now 
simply related to the consistent anomaly, the consistent anomaly
can be related to Berry's curvature too.
This is done by improving the adiabatic (section 7) 
phase choice so as to simplify the lattice 
Wess-Zumino functional, without loosing the linearity in the
gauge transform variable. The new form of the Wess-Zumino
functional involves 
Berry's curvature under the disguise of the expectation value 
of a nontrivial commutator of currents, a Schwinger term 
(section 8). 
After this first phase redefinition the road is open to take another
step of improvement, this time reducing the lattice Wess-Zumino term to
a minimal form which is a direct counterpart of the continuum form.
The second step is taken in detail in two dimensions 
(section 9). This part is
entirely technical and generalizations to higher dimensions than
2 can be tedious. Some of the tedium has been removed by recent work,
as shall be mentioned where appropriate (section 10). After that 
the lattice effective action for an anomalous
theory is briefly discussed in the context of its 
relation to the Preskill-Wise work (section 11). Towards
the end of the paper, the single particle version of the adiabatic
phase choice is worked out (section 12); this is used to show that the
good symmetry properties of the BW phase choice [\ovlapb] also hold for the
adiabatic phase choice made in this paper (section 13). After some brief
comments about the nonabelian case (section 14) 
the paper ends with a summary (section 15).

\vskip .5cm
\leftline{{\bf 2. Notations and Conventions}}
\vskip .5cm

As explained above, I work at infinite volume. No attempt will
be made to establish a thermodynamic limit in any rigorous way.
Still, at times it may be necessary to subtract some trivial
thermodynamic infinities. At those times I shall use $V$ to denote
the total number of sites on our $d$-dimensional hypercubic lattice.
The Dirac type operators will involve infinite matrices that
can be viewed as matrices of dimension $N\times N$ where 
$N=2^{d_2} V$. Here, $d_2 = {d\over 2}$ and the
dimension $d$ is always even. 
Dirac indices will be denoted by $\alpha, \beta$, etc.
Sites will be identified by $x,y$, etc. 

A fundamental set of fermionic creation/annihilation operators
are
$$\{ {\hat a}^\dagger_\alpha (x) , {\hat a}_\beta (y) \} = 
\delta_{\alpha\beta}
\delta_{xy}\eqno{(2.1)}$$
Operators in the associated Fock space will carry hats. Typically,
they are bilinear in the ${\hat a}$'s, of
the form ${\hat X} = {\hat a}^\dagger X {\hat a}$. 
Here, suppressed indices are
summed over. $X$ is the kernel of ${\hat X}$. 

I shall make frequent use of forward and backward directional finite
difference operators:
$$
\eqalign{\nabla_\mu^{x+} f(x) = &f(x+\mu ) -f (x)\cr
\nabla_\mu^{x-} f(x) = &f(x)-f(x-\mu )}\eqno{(2.2)}$$
$\mu ,\nu,$ etc. denote directions on the lattice. The site superscript
on $\nabla$ identifies the site index on which the finite
difference is taken. 

The noncompact vector potential will be denoted by $A$ in abbreviated
form. $A_\mu (x)$ is an unbounded real number 
associated with the link going from the site $x$ into the positive 
$\mu$ direction. A gauge transformation is defined by a set
of real numbers $\alpha(x)$ associated with sites on the lattice.
Under a gauge transformation we have:
$$
A_\mu (x) \to A_\mu (x) + \nabla_\mu^{x+} \alpha(x) \equiv A^{(\alpha)} 
(x)\eqno{2.3)}$$
The field strength associated with a plaquette starting at site $x$,
going one lattice spacing in the positive 
$\mu$ direction followed by another
step in the positive $\nu$ direction will be denoted by $F_{\mu\nu}(x)$.
$$
F_{\mu\nu} (x) = \nabla_\mu^{x+} A_\nu (x) - \nabla_\nu^{x+} A_\mu (x)
\eqno{(2.4)}$$
The pure gauge action is gauge fixed, and to avoid unnecessary 
infrared complications the photon is given a mass:
$$
S_{\rm pure ~gauge} =
{1\over {4e^2}} \sum_x F_{\mu\nu}^2  (x)
+{{m^2_\gamma}\over 2} \sum_x A_\mu^2 (x) +
{1\over {2\xi}} \sum_x [ \nabla_\mu^{x-} A_\mu (x) ]^2\eqno{(2.5)}$$
This makes the path integral over $A$ well defined for each momentum
mode. In the above formula I made standard implicit assumptions 
about contracting $\mu$ and/or $\nu$
indices. $e$ is the (unquantized) coupling
constant. The pure gauge theory is Gaussian. When matter is
coupled in a gauge invariant way, the gauge dependence of all
1PI correlation functions can be isolated in closed form
because the mass and gauge fixing terms are quadratic and
$A$ is not an angular field variable. 

In order to couple the gauge degrees of freedom to a fermion of
charge $q$ we introduce the 
unitary link variables $U_\mu (x)$, defined by
$$
U_\mu (x;q) = e^{iqA_\mu (x)}\eqno{(2.6)}$$
Note that the link variables are not fundamental, the fundamental
field is $A$. 
If we have a single fermion, its charge $q$ can be 
absorbed into a redefinition
of $e, m_\gamma,\xi$. When we have several fermions we shall absorb
in this way the charge highest in absolute value, $|q|$. Therefore,
when we deal with any fermion individually its charge 
$q$ obeys $|q|\le 1$.
In the following we only write equations for the highest charge
fermion, picked to have $q=1$; $U_\mu (x;1)\equiv U_\mu (x)$. 
To deal with a charge $q$ fermion
one simply has to replace $A$ by $qA$ everywhere in the context
of that fermion.  

Out of the link matrices $U$ we construct the
directional parallel transporters $T_\mu$, 
which act both on the site index and on 
the group index of fermions $\psi$:
$$
T_\mu (A) (\psi ) (x) = U_\mu (x) \psi (x+\hat\mu)\eqno{(2.7)}
$$

Euclidean Dirac matrices are denoted by $\gamma_\mu$,  
act only on spinorial indices, and are defined as usual: $\{\gamma_\mu ,
\gamma_\nu \} =2\delta_{\mu\nu}$. 
The chirality matrix $\gamma_{d+1} = i^{d_2} \gamma_1 \gamma_2 ...
\gamma_d$ anticommutes with all $\gamma_\mu$, obeys $\gamma_{d+1}^2 =1$, 
and is hermitian. 

Gauge transformations act 
by pointwise multiplication by 
$e^{i\alpha (x)}$. 
$$ 
(G(\alpha)\psi)(x) =e^{i\alpha (x)}\psi(x)\eqno{(2.8)}$$
The $T_\mu$ matrices are ``gauge covariant'',
$$
G(\alpha) T_\mu (A) G^\dagger (\alpha ) = T_\mu (A^{(\alpha )} ) 
\eqno{(2.9)}
$$

The continuum massive Dirac operator has many lattice 
analogues, all obeying full hypercubic symmetry. Among
these lattice matrices, the Wilson Dirac matrix,
$D_W (A)$ is the sparsest. $D_W (A)$ can be written as:
$$
D_W (A) = m+\sum_\mu(1- V_\mu );~~~V_\mu=
{{1-\gamma_\mu}\over 2} T_\mu +{{1+\gamma_\mu}\over 2} T_\mu^\dagger 
\eqno{(2.10)}
$$
One easily checks that $V_\mu^\dagger V_\mu =1$ which tells us that
$D_W (A)$ is bounded. 
The matrix $H_W (A)=\gamma_{d+1} D_W (A) $ is easily
seen to be hermitian.
The parameter $m$ is fixed at some value close to $-1$.   

Most of the time the $A$ configurations are restricted by 
requiring 
$$
|F_{\mu\nu} (x) | \le \eta\eqno{(2.11)}$$
for every plaquette. If this bound is obeyed for the fields
felt by the fermion of charge one, it evidently
is also obeyed by the fields seen by the charge $q$ fermions. 

Pick some number $\eta$ which obeys [\mybounds]
$$
0< \eta < {{1-(1+m)^2}\over {(1+{\sqrt{2}\over 2})d(d-1)}}
\eqno{(2.12)}$$
Clearly, one needs $|1+m|<1$. 
Since $|\sin(\theta ) | \le |\theta|$ for any $\theta$,
the matrix $H_W (A)$ falls in the class analyzed in 
reference [\mybounds]. As a result, the lowest eigenvalue
of $H_W^2(A)$, $\lambda_{\rm min} (A)$, obeys the following bound:
$$
\left [ \lambda_{\rm min} (A)\right ]^{1\over 2}\geq
\left [ 1 - \eta (1+{\sqrt{2}\over 2}) d(d-1) \right ]^{1\over 2} -|1+m| 
>0 \eqno{(2.13)}$$
Therefore, on the restricted space of configurations characterized
by (2.11) $H_W(A)$
never has a zero eigenstate. By a deformation argument this
proves that $H_W (A)$ has the same number of negative eigenvalues
as the free case $H_W (0)$. This number, $N_v$, is half the total
dimension $N$. Although both $N_v$ and $N$ are infinite, for 
an unrestricted field $A$ it would have been conceivable 
that $N_v - {1\over 2} N$ be some finite integer. This
integer is a meaningful quantity by deformation
arguments. The restriction on the field strength makes 
the integer vanish for all restricted $A$'s.
The space of restricted $A$'s is contractible to $A=0$
because if $A$ is in the restricted set, so is $tA$ where $t$
is between zero and one. 

Over the restricted set of $A$'s the sign function of $H_W(A)$,
$\epsilon (A)$, is well defined and local\footnote{${}^{f_2}$}{
The Ginsparg-Wilson relation [\gw] is equivalent to $\epsilon^2 (A)=1$.}. 
Below I define a Hamiltonian acting on the Fock space generated by polynomials
in ${\hat a}^\dagger$ acting on a vacuum annihilated
by all ${\hat a}$:
$$
{\hat H}(A)={\hat a}^\dagger \epsilon (A) {\hat a} + N_v ,~~~~\epsilon(A) 
={\rm sign} (H_W (A)) \eqno{(2.14)}$$
The role of the additive constant is to assure that the ground
state energy is zero. Another important operator in the Fock space
is the local charge:
$${\hat n}(x) = {1\over 2} 
[a^\dagger (x) a(x) - a(x)a^\dagger (x)]\eqno{(2.15)}$$
An additive constant was chosen so that 
the vacuum have zero total charge. The total charge operator is
$$
{\hat N} =\sum_x {\hat n}(x)\eqno{(2.16)}$$
The states in the Fock space will be denoted by a Dirac bra-ket
notation. The ground state has zero energy and zero total charge:
$$
{\hat H}(A)| v(A) \rangle =0,~~~~{\hat N} |v(A) \rangle =0
\eqno{(2.17)}$$

Under a gauge transformation $\epsilon(A)$ is gauge covariant,
inheriting this property from the parallel transporters, $T_\mu (A)$.
In the Fock space, gauge transformations are represented by
$$
{\hat G} (\alpha ) = e^{i\sum_x \alpha (x) {\hat n}(x)}\eqno{(2.18)}$$
The covariance of $\epsilon(A)$ implies:
$$
{\hat G} (\alpha ) {\hat H}(A) {\hat G}^\dagger (\alpha ) = {\hat H} 
(A^{(\alpha )} )\eqno{(2.19)}$$

The ground state of ${\hat H}(A)$ is obtained by occupying all negative
energy eigenstates of $\epsilon (A)$. The ground state is nondegenerate.
All single particle-hole excitations are degenerate, and have energy
equal to $2$. All energy eigenvalues are $A$-independent. 

Replacing $\epsilon (A)$ above by $\gamma_{d+1}$ and denoting the associated
Hamiltonian by ${\hat H}^\prime$ defines a reference ground state
$$
{\hat H}^\prime |v^\prime \rangle = 0\eqno{(2.20)}$$
Since $\gamma_{d+1}$ is diagonal in site space 
$$
{\hat G} (\alpha ) |v^\prime \rangle =|v^\prime \rangle\eqno{(2.21)}$$
if we require $\sum_x \alpha (x) =0$. 
Moreover, for a constant gauge transformation, where $\alpha(x)$
is independent of $x$, equation (2.17) still applies
because exactly half of the total number of states is filled in
the reference ground state,
just as in $|v(A)\rangle$. 
Therefore, equation (2.21) holds for all $\alpha$.

The overlap provides expressions for the fermion determinant and for
all fermionic correlation functions [\ovlapb]. 
Throughout this paper we shall only need an 
explicit formula for the determinant:
$$
\langle v^\prime | v(A)\rangle \eqno{(2.22)}$$
The absolute value of the determinant is
gauge invariant. However, the phase of $| v(A)\rangle$ has still
not been defined and whatever definition one chooses it
is implausible that the resulting phase of the overlap would
turn out gauge invariant. This paper is about how to define the
phase so that, on the lattice, before any limits are taken,
gauge invariance be violated by  an amount not larger than in continuum
formulations.

As far as the fermionic correlation functions go, 
all we need to know about them is that
by themselves they are gauge covariant; thus, any gauge breaking
effects reside in the fermionic determinant. In short, if
a gauge invariant phase choice is found the entire fermionic
sector is gauge covariant and the entire quantization procedure
of the chiral theory proceeds just as in textbook QED.

\vskip .5cm

\leftline{{\bf 3. Adiabatic phase definition}}
\vskip .5cm

There are many possibilities to make an adiabatic phase choice: 
one connects the Hamiltonian one is working with to a standard
reference Hamiltonian by a slow time 
evolution and makes a standard choice for
the phases of the eigenstates of the standard Hamiltonian. 

The earliest suggestion to employ an adiabatic phase choice for
the overlap was made in [\ovlapb]. Any adiabatic
phase choice involves an evolution law,
so would require some integration. 
This is not easy to implement numerically
and therefore previous work on the overlap almost exclusively
employed the so called BW phase choice. This phase choice
will be defined later on when its connection to the present
phase choice will be worked out. It is easier to work with the
BW phase choice numerically. Also, the BW phase choice is amenable
to perturbation theory. The adiabatic phase choice though, has
nicer properties and is more geometrical. The $BW$ phase choice
also has a geometrical interpretation, but it seems less useful
in the gauge theory context. 

An adiabatic phase choice
was also suggested in the past 
by S. Randjbar-Daemi and J. Strathdee [\daemi]. 
These authors used traditional, exponential in time, adiabatic 
turn on of the entire interaction piece of the
Lagrangian and showed that anomalies are reproduced
in perturbation theory with their phase choice. It was left unclear
whether this adiabatic phase choice preserved 
on the lattice as large a set of 
symmetries as the BW phase choice did. But, their work made it
quite evident that an adiabatic phase choice was a possible
alternative to the BW phase choice and is one of the main
motivations for this paper. 

My choice of adiabatic phase is different from that
of  S. Randjbar-Daemi and J. Strathdee 
in that I choose linear interpolation of gauge fields: The gauge fields
$A_\mu (x)$ appear as parameters in the Hamiltonian. They are
replaced by time dependent gauge fields $A_\mu (x,t) = t A_\mu (x)$. 
The time variable $t$ is taken to vary between zero and one, but 
the Hamiltonian is assumed multiplied by a large number $T$, so that
the evolution 
$$
i|{{d  v (t;A) }\over {dt}}\rangle_T = T{\hat H (tA ) } 
| v (t;A) \rangle_T \eqno{(3.1)}$$
is almost adiabatic. At $t=0$ 
$$
|v (t=0;A) \rangle_T  = | v_0 \rangle \eqno{(3.2)}$$
where $| v_0 \rangle$ is the ground state of $\hat H (0)$ with 
a specific phase choice. Multiplying $| v_0 \rangle$ by
a pure phase makes all subsequent states in the evolution
change by the same phase, which becomes an $A$-independent, 
immaterial constant. 

$\hat H (A)$ has zero ground state energy
at all $A$. In the limit of large $T$ the adiabatic theorem
tells us that for those gauge fields that the ground state 
of $\hat H (A)$ is separated by a gap from the excited state, 
the following limit exists and is approached with corrections
that go as ${1\over T}$ [\kato]:
$$
\lim_{T\to\infty} |v (t;A) \rangle_T = 
|v (tA ) \rangle , ~~t>0\eqno{(3.3)}$$
The states $|v(A)\rangle$ play an important role in what
follows; they are uniquely fixed by two conditions:
$$\eqalign{
{\hat H}& (A )|v (A ) \rangle =0\cr
\langle & v(tA )|{ {dv (tA )} \over{dt}}  \rangle =0} \eqno{(3.4)}$$
The states $ |v (A )\rangle$ depend smoothly on $A$. 
With their help the overlap
$$
\langle v^\prime | v(A)\rangle\eqno{(3.5)}$$
is completely defined. The phase choice is fixed by the second line
in equation (3.4). 

There is another way to view the adiabatic phase choice: The
state $|v(A)\rangle$ can be viewed as a complex scalar field
in a $CP({\cal N})$ model with very large ${\cal N}$ and with
the gauge fields $A$ as base space. Berry's
connection is the $U(1)$ gauge field a physicist would naturally 
associate with a $CP({\cal N})$ model. When the $U(1)$ in
a $CP({\cal N})$ model is gauged one writes an action that does
not depend on ``local'' (this means $A$-dependent) phase transformations
of the states $|v(A)\rangle$. This $U(1)$ gauge field over the space
of $A$'s changes by a $U(1)$ gauge transformation when the state
$|v(A)\rangle$ is multiplied by a phase. Therefore, 
local phase independence is achieved by writing a Lagrangian
in terms of the abelian field strength associated with the
gauge field. Since the gauge field is Berry's connection 
${\cal A}_{\mu x} (A)$, the field  
strength ${\cal F}_{\mu x,\nu y} (A)$ is Berry's curvature. 
$$
{\cal A}_{\mu x} (A) = \langle v(A) | {{\partial v(A)}\over {\partial
A_{\mu} (x)}} \rangle\eqno{(3.6)}
$$
A phase choice for $|v(A)\rangle$ amounts to fixing the gauge in
this $U(1)$ gauge theory over $A$-space. The BW phase choice more
or less corresponds to a ``unitary gauge'' because it fixes the
phase of the component of $|v(A)\rangle$ in a given direction in
the Hilbert space. The adiabatic phase choice we adopt in this 
paper corresponds to the Fock-Schwinger gauge choice:
$$
\sum_{\mu x} A_\mu (x) {\cal A}_{\mu x} (A) =0
\eqno{(3.7)}$$
In (3.7) I chose to make explicit also the summation over $\mu$, just
to stress the similarity to the more familiar form of the Fock-Schwinger
gauge in ordinary continuum QED, $\sum_\mu x_\mu A_\mu (x)=0$. 
Equation (3.7) is trivially equivalent to the second line in equation
(3.4). The Fock-Schwinger gauge is particularly appropriate for
abelian gauge theories on contractible spaces, which is the case here.

\vskip .5cm
\leftline{{\bf 4. 
The adiabatic phase choice produces a linear Wess-Zumino action}}
\vskip .5cm

The goal now is to derive the relation between two states corresponding
to backgrounds differing by a gauge transformation $\alpha$:
$$
A^{(\alpha)}_\mu (x) = A_\mu (x) + \alpha (x+\mu ) - \alpha (x)
\eqno{(4.1)}$$
Recall that the Hamiltonian is gauge covariant 
and that there
is no degeneracy in the ground state. Therefore:
$$
|v (A^{(\alpha )}) \rangle = e^{i\Phi (\alpha , A ) } 
G (\alpha ) |v (A)\rangle \eqno{(4.2)}$$
This defines the Wess-Zumino action $\Phi (\alpha , A )$. 
To calculate $\Phi (\alpha , A )$ the adiabatic phase 
definition of equation (3.4) must be used. Therefore,
the ``time'' parameter $t$ has to be reintroduced:

Fix $\alpha$ and $A$. For all $t$,
$$
|v ( (tA)^{(t\alpha )}) \rangle = e^{i\Phi (t\alpha , tA ) }
G (t\alpha ) |v (tA)\rangle\eqno{(4.3)}$$
with $\Phi (t\alpha , tA )=0$ at $t=0$. The
gauge transform acts linearly:
$$
(tA)^{(t\alpha )}_\mu (x) =tA_\mu (x) +t(\alpha(x+\mu)-\alpha(x)) 
= t A^{(\alpha)}_\mu (x)\eqno{(4.4)}$$
This relation becomes more complicated in the nonabelian case. 
That it holds here is the basic reason for the linearity
of $\Phi(\alpha ,A)$ in $\alpha$.

For brevity, let us temporarily 
denote $\Phi (t\alpha , tA )=\varphi (t)$. 
Now, take a time derivative of 
equation (4.3), and after that multiply the resulting
equation from the left by $\langle v(t A^{(\alpha )} )|
=\langle v((t A)^{(t\alpha )} )|$. On the left there is one term
and on the right there are three terms, one for each time dependent factor
on the right hand side of equation (4.3).
$$
\langle v(tA^{(\alpha )} ) | {{dv(tA^{(\alpha )})}\over {dt}} \rangle =
i{{d\varphi}\over {dt}}
+i\sum_x \alpha (x) \langle v(t A) | {\hat n} (x) 
|v(tA) \rangle + \langle v(tA)| {{dv(tA)}\over {dt}}
\rangle \eqno{(4.5)}
$$
The adiabatic condition makes the terms containing
time derivatives acting on states vanish, leading to
$$
{{d\varphi}\over {dt}}=-\sum_x \alpha (x) \langle
 v (tA)|{\hat n}(x) | v(tA )\rangle \eqno{(4.6)}
$$
The main result is quite simple:
$$
\Phi (\alpha , A) = -\sum_x \alpha (x) \int_0^1 dt
\langle v (tA)|{\hat n}(x) | v(tA )\rangle \eqno{(4.7)}$$
The important features of this formula are that it is linear
in the gauge degrees of freedom $\alpha (x)$ and that the factor
multiplying the $\alpha(x)$ is a gauge invariant function of $A$. 
Moreover, 
since $\sum_x {\hat n}(x) = {\hat N}$ is the total fermion number
operator defined so that 
$$
{\hat N} |v (A )\rangle =0,\eqno{(4.8)}$$
we have the identity
$$
\sum_x \langle v (A)|{\hat n}(x) | v(A )\rangle =0\eqno{(4.9)}$$
for all $A$. Although (4.9) has an infinite sum over sites,
there is nothing ``formal'' about it because
the matrix element vanishes at large $x$. This is so because the
action forces $A_\mu (x)$ to go to zero as $x$ goes to infinity
and the state $|v(A) \rangle$ comes from a massive field theory,
ensuring that the matrix element is a local functional of $A$. 

Since the overlap is the regulated chiral fermion determinant,
the functional $\Phi(\alpha ,A)$ indeed deserves to be
viewed as the regulated abelian 
Wess-Zumino action describing the
gauge dependence of the lattice fermion determinant. 
$$
\langle v^\prime | v(A^{(\alpha)})\rangle =
e^{i\Phi(\alpha ,A)}\langle v^\prime | v(A)\rangle\eqno{(4.10)}$$
Here we used that ${\hat n}(x)|v^\prime\rangle = 0$ for all $x$.

The derivation of the main result
and its main properties was simple because the
underlying physics is simple: Think in terms of the Callan
Harvey setup, but adopt the overlap viewpoint of the coordinate
perpendicular to the domain wall as a time direction. ${\hat n} (x)$ 
becomes then the local charge operator and the total charge ${\hat N}$
is conserved. The state $|v(A)\rangle$ is built up slowly starting
from $|v_0 \rangle$ and gradually increasing $A$ to its final value.
The local charge density, $\langle v(A) | {\hat n}(x) |v (A)\rangle$
starts off at $A=0$, where it vanishes. The total charge $\sum_x
\langle v(A) | {\hat n}(x) |v (A)\rangle$ is conserved, so that
the single way local charges $\langle v(A) | {\hat n}(x) |v (A)\rangle$
build up is by flow of charge carrying currents moving
local charge from one place to another. The flow
of these currents during the slow buildup causes local charge to be
redistributed, making the
expectation value of ${\hat n} (x)$ nontrivially dependent on 
time and its final value on the final field $A$. 
The history of the slow
buildup of local charge tells us what phases
will be acquired when we change the background by a 
gauge transformation because the local charge is also
the generator of local phase transformations. This picture not
only explains the formula for $\Phi(\alpha , A)$, but also
indicates that the charge buildup is described
by local currents. This will play a role in what follows. 

Before proceeding let me remark that
the adiabatic phase choice makes the action induced by 
integrating out the fermions depend on the entire, single site, noncompact
vector potential $A_\mu (x)$, and not just on the vector
potential modulo some integer valued field, as would be the
case if all dependence went through $e^{iA_\mu (x)}$. In
this sense, the effective action induced by integrating
out the fermions is similar to the pure gauge action. 
However, it is only the phase choice that depends 
also on the $2\pi\times $ integer part of the vector
potential. The real part of the fermionic contribution 
to the action depends only on the $e^{iA_\mu (x)}$'s. 

It is worthwhile to stress that also the variables $\alpha (x)$
do not appear as angular variables in the above Wess-Zumino
action. In other words, $\Phi(\alpha ,A)$
does not change by an integral multiple of $2\pi$
when we shift $\alpha(x) \to \alpha(x) +2\pi z(x)$
with $z(x)\in Z$. This is intrinsically related to the 
linearity of $\Phi(\alpha ,A)$ in $\alpha$. For example,
the $BW$ phase convention keeps both $A_\mu (x)$ and $\alpha(x)$
at the status of angular variables. This is a source of difficulties,
see for example [\chiuone]. While ``unrolling'' $A_\mu (x)$
may seem a mere technicality, doing the same for the gauge transformation
parameter, on the lattice, is a bit more surprising:

A very well known view of anomalies, due to Fujikawa [\fuji], is that
the Wess-Zumino action is the Jacobian associated with the
change in fermion integration measure under the gauge transformation
$\psi_R (x) \to e^{i\alpha (x)} \psi(x)$. In particular
on a lattice this really means, for example, that a transformation
with $\alpha (x) = 2\pi z(x)$ does not do a thing. But, with
our adiabatic phase choice, we do get a nontrivial Wess-Zumino
action for such a gauge transformation! 
Hence, a literal realization of Fujikawa's interpretation
is ruled out in the overlap defined with the adiabatic phase choice. 
Still, much of the essence of Fujikawa's viewpoint is preserved:
The lack of gauge invariance is entirely contained in the 
Wess-Zumino term and there are no other sources of gauge breaking;
all this is just as it would be if the anomaly truly could be
viewed as a fermion integration measure effect. But, on the lattice, 
it seems that insisting on a ``measure'' terminology is 
inappropriate. 
In short, the Wess-Zumino action 
is the single source of gauge noninvariance because the overlap
preserves the slightly amended 
continuum Fujikawa viewpoint, namely that {\it all}
gauge violation can be viewed as if coming from a gauge field dependent 
fermionic integration ``measure''. Thus, if the Wess-Zumino action
vanishes, full lattice gauge invariance gets restored. Taking
the word ``measure'' to really mean measure however, is useless and
confusing.

In the continuum the imaginary part of the
logarithm of the chiral determinant is parity odd and purely
imaginary. It is also gauge dependent
and gives the anomaly in the so called consistent form.
In the abelian case the gauge dependence is exactly 
linear in the gauge transform
variables $\alpha$. The coefficient of $\alpha$
is gauge invariant and homogeneous in the field strength where the
degree of homogeneity 
is defined by the number of field strength factors required 
to saturate the $d$-dimensional antisymmetric epsilon symbol.

With the overlap and the adiabatic phase choice we preserve most
of the properties listed in the previous paragraph 
(symmetries ensuring parity oddness will be
dealt with later, in section 13) 
except that the coefficient of $\alpha$ is not
that simple. Had we used the BW phase choice instead, we would have
also lost the linearity in $\alpha$.

Let us now turn to establish some direct consequences of the linearity
of the lattice Wess-Zumino action with the adiabatic phase choice.

\vskip .5cm
\leftline{{\bf 5. Gauge invariance on the trivial gauge orbit}}
\vskip .5cm

The trivial orbit is given by gauge transforms of $A=0$. The main result
of the previous section implies that the dependence on $\alpha$ is now
$$
\Phi (\alpha , 0) = -\sum_x \alpha (x) \langle v(0)|{\hat n} (x)
\rangle |v (0)\rangle \eqno{(5.1)}$$
Because of translational invariance the matrix element 
$\langle v(0)|{\hat n} (x) \rangle |v (0)\rangle$ must be a constant
and this constant is zero since the total
charge of all $| (A)\rangle$ is zero. Thus, there is no gauge
dependence on the trivial orbit. This property was shown to be true
in 2 dimensions with the BW phase choice [\ovlapb]. With the adiabatic
phase choice we now see that it holds in any dimension.

Note that absence of dependence on the gauge degrees of freedom
on the trivial orbit holds independently of whether the theory
is anomalous or not. This is what one would expect, based on the
continuum, at infinite volume. However, in the past, people often
focused on models reduced to the 
trivial orbit and the failure to satisfactorily
eliminate the gauge dependence there (for example, recall the Yukawa/gauge
fixed approach [\shamir]) was taken 
as an indication of the difficulties associated
with lattice chiral fermions. It is now apparent that this line of thought
was in error, as probably many long suspected when observing that
difficulties were appearing even before a single fermion loop was
taken into account.

\vskip .5cm
\leftline{{\bf 6. Simple relation between consistent and covariant anomalies}}
\vskip .5cm

Since we are dealing with the abelian case the terminology in
the title is misleading:
both the consistent and the covariant anomaly are gauge invariant
in the continuum. This is also true on the lattice with the overlap
in the adiabatic phase choice. Nevertheless, there are two anomalies:
they differ by a prefactor that can be understood as coming from
imposing full Bose symmetry on the external legs of the
anomaly diagram in the consistent case. 

The lattice consistent anomaly is trivially read off the function $\Phi$:
$$
\triangle_{\rm consistent} (x;A) = -\int_0^1 dt \langle v(tA)|{\hat n}(x)
|v(tA)\rangle\eqno{(6.1)}$$
The covariant anomaly differs from the consistent anomaly by the
divergence of the Berry connection viewed as a current. To understand
this statement I need to review the definitions of the consistent
and covariant currents in the overlap context:

The nonlocal consistent current, given by definition by 
$$
{\cal J}_{\mu \ {\rm consistent}}
(x;A) = {{\partial \log \langle v^\prime | v (A) \rangle}\over
{\partial A_\mu (x)}}\eqno{(6.2)}$$
is naturally decomposed into a gauge invariant part and a local part:
$$
{\cal J}_{\mu \ {\rm consistent}} (x;A) = {1\over
{ \langle v^\prime | v(A) \rangle}}
\langle v^\prime |{{ \partial v (A) }\over
{\partial A_\mu (x)}}\rangle_\perp +
\langle v(A) | {{\partial v (A) }\over
{\partial A_\mu (x)}}\rangle \eqno{(6.3)}
$$
The first term is the gauge invariant (covariant in the nonabelian case)
${\cal J}_{\mu \ {\rm covariant}}(x;A)$ current 
and the second term is Berry's
connection ${\cal A}_{\mu x} (A)$. 
By definition $|{{ \partial v (A) }\over
{\partial A_\mu (x)}}\rangle_\perp = \left [ 1- |v(A)\rangle \langle v(A)|
\right ] |{{ \partial v (A) }\over
{\partial A_\mu (x)}}\rangle$. A more explicit formula for
${\cal J}_{\mu \ {\rm covariant}}(x;A)$ will be derived in section
8. Berry's connection is defined
only in terms of the state $|v(A)\rangle$ which is the ground state
of a massive system; this makes ${\cal A}_{\mu x} (A)$ a local 
functional of $A$. The consistent and covariant
currents depend on both states $|v^\prime \rangle$
and $|v(A)\rangle$ and are nonlocal functionals of the $A$; it is only the
current difference that is independent of $|v^\prime \rangle$
and therefore local. 

The consistent anomaly $\triangle_{\rm consistent} (x;A)$
quoted above is the divergence of the consistent current
$$
\triangle_{\rm consistent} (x;A)=-\nabla_\mu^{x-} {\cal J}_{\mu \ {\rm
consistent}}(x;A),\eqno{(6.4)}$$
and the covariant anomaly $\triangle_{\rm covariant} (x;A)$
quoted above is the divergence of the covariant current
$$
\triangle_{\rm covariant} (x;A)=-\nabla_\mu^{x-} {\cal J}_{\mu \ {\rm
covariant}}(x;A) \eqno{(6.5)}$$
Therefore, by definition, 
$$
\triangle_{\rm consistent} (x;A) - \triangle_{\rm covariant} (x;A) = -
\nabla^{x-}_\mu 
\langle v(A) | {{\partial v (A)}\over {\partial A_\mu (x) }}
\rangle\eqno{(6.6)}$$
We can explicitly evaluate the right hand side
by making $\alpha$ infinitesimal in the 
gauge transformation rule of the states $|v(A)\rangle$:
$$\eqalign{
\langle v(A)| v(A^{(\alpha )}) \rangle =1+&
e^{i\Phi (\alpha, A)} \langle v(A) | G(\alpha ) |v(A) \rangle =\cr
& i\sum_x \alpha(x) \int_0^1 tdt 
{{d\langle v(tA)|{\hat n}(x)| v(tA)\rangle}\over {dt}} +
O(\alpha^2 )}\eqno{(6.7)}$$
This implies:
$$
\triangle_{\rm covariant} (x;A) = -
\langle v(A)|{\hat n}(x)|v(A)\rangle\eqno{(6.8)}$$
One can directly verify quite easily 
that indeed the above is the divergence
of the covariant current as defined below equation (6.3) [\dubna].

That the relationship between the two anomalies 
comes out right in the continuum
limit is obvious if one accepts that the continuum limit of the
covariant anomaly is homogeneous in $A$. Then one has:
$$
\triangle_{\rm consistent} (x;A)=c\triangle_{\rm covariant} (x;A)
\eqno{(6.9)}$$
The degree of homogeneity is $d_2$ because one needs to saturate
the antisymmetric $d$ dimensional epsilon symbol by field strength factors. 
The integral over $t$ in the consistent anomaly is trivial. We obtain
a result better known from arguments based on the symmetry of a one
fermion loop diagram:
$$
c={1\over {1+d_2}}\equiv {1\over {1+{d\over 2}}}\eqno{(6.10)}$$
Establishing this relation, for example, with the BW phase choice is more
difficult. 

The main result here is that, even before the continuum limit
is taken, at finite lattice spacing, the difference between the
two anomalies is quite similar in structure to either anomaly:
$$
\triangle_{\rm consistent} (x;A) - \triangle_{\rm covariant} (x;A)
=\int_0^1 tdt {{d\langle v(tA)|{\hat n}(x)| v(tA)\rangle}\over {dt}}
\eqno{(6.11)}$$
From (6.11) we learn that in order to set the difference between
the consistent and covariant anomaly to zero, identically for all
$A$, we need $\langle v(A)|{\hat n}(x) |v(A)\rangle \equiv 0$.
But, if this is true all anomalies vanish. In short, making the
difference between the two anomalies vanish makes them vanish
individually. 

On the other hand, 
the difference
between the consistent and covariant anomalies is governed by Berry's
curvature, as explained in [\geom]. Berry's
curvature is a gauge invariant object at finite lattice spacing. 
It is a rank two antisymmetric tensor over the space of gauge fields $A$.
We conclude that it must be possible to improve the
adiabatic phase choice so as to make it explicit that 
anomalies are non-vanishing if and only if Berry's curvature
is nonzero. This is the objective of the next section.

\vskip .5cm
\leftline{{\bf 7. Improving the adiabatic phase choice}}
\vskip .5cm
To simplify further the 
Wess-Zumino lattice action 
we need to change the adiabatic phase choice.
After this simplification 
it will become evident that a further phase redefinition
exists for which the Wess-Zumino term 
vanishes altogether if (and only if) perturbative anomalies cancel.

The adiabatic phase choice obeys the symmetries first established
for the BW phase choice in [\ovlapb] (the proof will
be presented in section 13). Since the BW phase choice and the adiabatic phase
choice both obey the symmetries of [\ovlapb], so does their
difference, so one can view our final phase choice as coming from
the BW phase choice directly. 
But, it is easier to
do get there starting from the adiabatic phase choice.

To redefine the phase so that the Wess-Zumino action
$\Phi (\alpha ,A)$ simplifies as much as possible 
we need a better understanding of
the quantity $\langle v(A) | {\hat n }(x) | v(A) \rangle$. 
As discussed before, the basic
physics observation is that $\langle v(A) | {\hat n }(x) | v(A) \rangle$
is built up adiabatically from $0$ by the flow of currents.
Let us look for the current by calculating the time evolution
of the local charge:
$$\eqalign{
{{d \langle v(tA) | {\hat n }(x) | v(tA) \rangle}\over {dt}}=&
2\Re [ \langle v(A)| {\hat n}(x) | {{d v(tA )}\over {dt}} \rangle ]
=\cr 2\Re &\left [ 
\sum_n \langle v(tA) | {\hat n}(x) {\hat H}(tA)| v_n (tA) \rangle 
{1\over{E_n (tA)}}
\langle v_n (tA ) | {{d v(tA )}\over {dt}} \rangle \right ]}
\eqno{(7.1)}$$
The sum over $n$ extends over all single particle-hole excited
states above the ground state. The energies $E_n (A)$ are
all positive and equal to 2. 
Since the ground state energy has been chosen to be zero we can
replace ${\hat n}(x) {\hat H}(tA)$ by the commutator 
$[{\hat n}(x), {\hat H}(tA)]$. This is a commutator between bilinears
in creation-annihilation operators, so it is completely determined
by the commutators of the matrix kernels. This is a very familiar
exercise and we immediately conclude that
$$
[{\hat n}(x), {\hat H}(A)]= i \nabla_\mu^{x-} {\hat J}_\mu (x;A)
\eqno{(7.2)}$$
where the kernel of the current operator ${\hat J}_\mu (x;A) =
{\hat J}_\mu^\dagger (x;A)$ is
local in $A_\mu (x)$. So, we obtain
$$
\langle v(A) | {\hat n }(x) | v(A) \rangle = i \nabla_\mu^{x-}\int_0^1 dt
{1\over 2}
\left [ \langle v(tA) | {\hat J}_\mu (x;tA) 
| {{dv(tA)}\over {dt}}\rangle -c.c.\right ]\equiv 
\nabla_\mu^{x-} j_\mu (x; A)\eqno{(7.3)}$$
Although the left hand side is gauge invariant, the ``current''
$j_\mu (x; A)$ is not, as we shall see below. Still, under a gauge
transformation, $j_\mu (x; A)$ transforms linearly. $j_\mu (x; A)$ 
is explicitly constructed in the massive theory 
and therefore is a local gauge functional 
of the gauge fields. This means that its value at $x$ depends
on values of $A_\mu (y)$ only exponentially weakly as $|y-x|\to\infty$.

The current is used to redefine the phases of the adiabatic states by
$$
|v(A) \rangle \to |v^{\rm new}(A)\rangle =
\exp\left [-i\int_0^1 dt A_\mu (x) j_\mu (x;tA)
\right ] |v(A) \rangle \eqno{(7.4)}$$
By design, the contribution 
from taking a gauge variation of the factor 
$A_\mu (x)$ multiplying the current in the first term
on the right hand side of equation (7.4) 
exactly cancels the adiabatic $e^{i\Phi(\alpha ,A)} $ pre-factor 
one gets from gauge transforming the state. 
$$\eqalign{&
|v^{\rm new} (A^{(\alpha)} ) \rangle = \cr
&e^{i\sum_x \alpha(x) 
\int_0^1 dt \nabla_\mu^{x-} j_\mu (x; tA)} e^{i\Phi(\alpha, A)}
e^{-i\sum_x A_\mu (x) \int_0^1 dt [j_\mu (x;tA^{(\alpha)})-
j_\mu (x; tA)]}
{\hat G}(\alpha ) | v^{\rm new} (A)\rangle \cr &= 
e^{-i\sum_x A_\mu (x) \int_0^1 dt [j_\mu (x;tA^{(\alpha)})-
j_\mu (x; tA)]}
{\hat G(\alpha )}| v^{\rm new} (A)\rangle }\eqno{(7.5)}$$

Taking into account ${\hat G}(\alpha) |v^\prime \rangle
=|v^\prime \rangle$ for all $\alpha$ we obtain the new
lattice Wess-Zumino action:
$$
\langle v^\prime | v^{\rm new} (A^{(\alpha)}) \rangle =
e^{-i\sum_x A_\mu (x) \int_0^1 dt [j_\mu (x;tA^{(\alpha)})-
j_\mu (x; tA)]}
\langle v^\prime | v^{\rm new} (A) \rangle\eqno{(7.6)}$$
What is happening reflects the well known ambiguity of the Wess-Zumino
orbit action: namely it can be altered by gauge transforms of 
arbitrary local functionals of the gauge field. The new lattice form of
the Wess-Zumino action is determined by the gauge transformation
properties of the current $j_\mu (x;A)$.

Recall the formula for the current $j_\mu (x;A)$:
$$
j_\mu (x; A) = i\int_0^1 dt
{1\over 2}
\left [ \langle v(tA) | {\hat J}_\mu (x;tA) 
| {{dv(tA)}\over {dt}}\rangle -c.c.\right ]\eqno{(7.7)}$$
The current
is not gauge invariant because of the derivative acting on the
state. When we insert the known gauge dependence of the states, and use the
gauge covariance of the operator ${\hat J}_\mu (x;tA)$
we get two contributions, one from ${\hat G}(t\alpha)$ and the other from
$e^{i\Phi (t\alpha , tA )}$:
$$ \eqalign{
&j_\mu (x;A^{(\alpha )}) - j_\mu (x; A)  = -
{1\over 2}\sum_y \alpha (y) \int_0^1 dt \left [ \langle v(tA) | {\hat J}
(x;tA) {\hat n}(y) |v(tA)\rangle + c.c.\right ] +\cr
& {1\over 2} \int_0^1 dt \sum_y \alpha(y) \langle v(tA)| {\hat n} (y) | v(tA ) \rangle
\left [ \langle v(tA)| {\hat J}_\mu (x;tA)) | v(tA ) \rangle + c.c \right ]}
\eqno{(7.8)}$$
In the first term above, 
${\hat n} (y)$ will either produce the ground state back,
whose energy is zero, 
or create a single particle-hole excitation of energy 2. 
The term containing the ground state will cancel against the
second term above, so we are left only with single pair excited
states. Therefore we can
insert ${\hat H}(tA)$ into the matrix element of the first
term, divide by 2 outside and remove the second term. Further,
the product ${\hat H}(tA){\hat n}(y)$ can be replaced by a commutator
since the ground state energy is zero. 
Using the definition of the operator ${\hat J}_\mu (x;A)$, we end
up with
$$ \eqalign{j_\mu (x;A^{(\alpha )}) - &j_\mu (x; A)=
{i\over 4} \sum_y  \left [
\int_0^1 dt \langle v(tA)| [ {\hat J}_\mu (x;tA), {\hat J}_\nu (y;tA)]|v(tA)
\rangle - c.c. \right ]  \nabla_\nu^{y+} \alpha (y) \cr
=&{i\over 2} \sum_y  \left [
\int_0^1 dt \langle v(tA)| [ {\hat J}_\mu (x;tA), {\hat J}_\nu (y;tA)]|v(tA)
\rangle \right ] \nabla_\nu^{y+} \alpha (y) \cr
}\eqno{(7.9)}$$
We are thus led to introduce the ``Schwinger term'', $S_{\mu\nu} (x,y;A)$:
$$
S_{\mu\nu} (x,y;A)= \langle v(A)| [ {\hat J}_\mu (x;A), {\hat J}_\nu
(y;A)]|v(A)\rangle\eqno{(7.10)}
$$
All the information about the gauge dependence of the current is
contained in the Schwinger term. The Schwinger term is closely related to 
Berry's curvature introduced in the overlap
context in [\geom]. Thus 
our objective from the previous section has been realized. 
The precise relation between the Schwinger term and Berry's
curvature is interesting in its own right and shall be worked out
in the next section. 

We thus learn that the new Wess-Zumino action is:
$$
i\Phi^{\rm new} (\alpha ,A) =-
{1\over 2} \sum_{x,y} A_\mu (x) [ \int_0^1 dt
\nabla_\nu^{y-} S_{\mu\nu}(x,y;tA) ] \alpha (y) \eqno{(7.11)}
$$
Note that $S_{\mu\nu} (x,y;A)$ is purely imaginary and antisymmetric
under simultaneous switch of $\mu$ with $\nu$ and $x$ with $y$. 
$S_{\mu\nu} (x,y;A)$ is bilocal, gauge invariant 
and obeys the rather restrictive identity:
$$
\nabla_\mu^{x-}\nabla_\nu^{y-} S_{\mu\nu} (x,y;A)\equiv 0\eqno{(7.12)}$$
This identity is easy to prove:
$$\eqalign{
\nabla_\mu^{x-}\nabla_\nu^{y-} S_{\mu\nu} (x,y;A) &\propto
\langle v(A) | \left [ {\hat n} (x) {\hat H}^2 (A) {\hat n} (y) - 
{\hat n} (y) {\hat H}^2 (A) {\hat n} (x)\right ] | v(A)\rangle  \cr &\propto
\langle v(A) | [{\hat n} (x),{\hat n} (y)] | v(A)\rangle }
\eqno{(7.13)}$$
The abelian structure of the group is the essential. By ``bilocal'' I
mean that $S_{\mu\nu} (x,y;A)$ approaches zero as $|x-y|\to \infty$, 
exponentially, with a decay that is bounded away from zero
uniformly in $A$ and $x$. Moreover, the dependence on $A_\mu (z)$
decreases exponentially with $z$ 
as both $|x-z|$ and $|y-z|$ go to infinity.

The identity in equation (7.12)
establishes that the coefficient of $\alpha$ in equation (7.11)
indeed is gauge invariant. To make this explicit we would like 
to replace the $A$ factor by a field strength factor $F$. This is indeed
possible as we shall see in equation (9.4).  

\vskip .5cm
\leftline{{\bf 8. Schwinger term, Berry's curvature and covariant current}}
\vskip .5cm
Berry's curvature ${\cal F}$ 
is defined as the curl over the space of $A$'s
of Berry's connection ${\cal A}$:
$$
{\cal A}_{\mu x} (A) = \langle v(A) | {{\partial v(A)}\over
{\partial A_\mu (x) }} \rangle \eqno{(8.1)}$$
$$
{\cal F}_{\mu x, \nu y} (A) = {{\partial {\cal A}_{\mu x} (A)}
\over {\partial A_\nu (y) }} - {{\partial {\cal A}_{\nu y} (A)}
\over {\partial A_\mu (x) }}=
\langle {{\partial v(A)}\over {\partial A_\mu (x)}} | 
{{\partial v(A)}\over {\partial A_\nu (y) }} \rangle -
\langle {{\partial v(A)}\over {\partial A_\nu (y)}} | 
{{\partial v(A)}\over {\partial A_\mu (x) }} \rangle
\eqno{(8.2)}
$$

A connection is sought between $ {\cal F}_{\mu x, \nu y} (A)$
and $S_{\mu\nu} (x,y; A)$. For this we need a formula for the current
operators ${\hat J}_\mu (x;A)$ which were defined by requiring locality
and 
$$
[{\hat n} (x) , {\hat H} (A)] = i\nabla_\mu^{x-} {\hat J}_\mu (x;A)
\eqno{(8.3)}$$
Expanding
$$
e^{i\sum_x \alpha (x) {\hat n} (x)} {\hat H} (A)
e^{-i\sum_x \alpha (x) {\hat n} (x)} = {\hat H}(A^{(\alpha)})
\eqno{(8.4)}$$
to linear order in $\alpha$ we immediately learn that a possible choice
for the current is:
$$
{\hat J}_\mu (x;A) = -{{\partial {\hat H} (A)}
\over {\partial A_\mu (x) }}\eqno{(8.5)}$$
Since ${\hat H} (A) | v(A) \rangle =0$ we have:
$$
{{\partial {\hat H} (A)} \over {\partial A_\mu (x)}} |v(A) \rangle =
-{\hat J}_\mu (x;A)|v(A) \rangle = -
{\hat H} (A) | {{\partial v(A)} \over {\partial A_\mu (x) } } \rangle
\eqno{(8.6)}$$
Hence,
$$\eqalign{
S_{\mu\nu} (x,y;A)=& \langle v(A)| [ {\hat J}_\mu (x;A), {\hat
J}_\nu
(y;A)]|v(A)\rangle = \cr
&\langle {{\partial v(A)}\over {\partial A_\mu (x)}} | {\hat H}^2 (A) |
{{\partial v(A)}\over {\partial A_\nu (y) }}\rangle  -
\langle {{\partial v(A)}\over {\partial A_\nu (y)}} | {\hat H}^2 (A) |
{{\partial v(A)}\over {\partial A_\mu (x) }}\rangle }
\eqno{(8.7)}$$
The states $|{{\partial v(A)}\over {\partial A_\mu (x) }}\rangle $ are a
linear combination of the ground state and single particle-hole
excitations of energy 2. For the excited states we can replace
${\hat H}^2 (A)$ by 4. For the ground state we get no contribution.
But, the ground state does not contribute as an intermediate state
to the inner products defining ${\cal F}$ either because of antisymmetry
and the purely imaginary character of ${\cal A}$. Hence,
$$
S_{\mu\nu} (x,y;A)=4{\cal F}_{\mu x, \nu y} (A)
\eqno{(8.8)}$$
Thus, up to a trivial proportionality factor the Schwinger term
is the same as Berry's curvature. 

Let us now derive a more physical formula for the covariant current
functional, ${\cal J}_{\mu \ {\rm covariant}}(x;A)$. In the abelian
case this gauge covariant current is actually gauge invariant. This
was proven in the general case (abelian and nonabelian) in ref [\geom].
We wish to derive an expression in which the 
gauge invariance becomes explicit. The state
$|{{ \partial v (A) }\over
{\partial A_\mu (x)}}\rangle_\perp$ is the same as the state
$|{{ \partial v (A) }\over
{\partial A_\mu (x)}}\rangle$ only the component in the direction
of $|v(A)\rangle$ is removed. The state $|{{ \partial v (A) }\over
{\partial A_\mu (x)}}\rangle$ contains one component in the direction
of the ground state and all other components are single particle hole
excited states of energy $2$. Therefore:
$$
|{{ \partial v (A) }\over
{\partial A_\mu (x)}}\rangle_\perp = {1\over 2}
{\hat H}(A) |{{ \partial v (A) }\over
{\partial A_\mu (x)}}\rangle\eqno{(8.9)}
$$
Using equation (8.6), we can now write the covariant current as:
$$
{\cal J}_{\mu \ {\rm covariant}}(x;A)={1\over 2}
{{\langle v^\prime | {\hat J}_\mu (x; A) |v(A)\rangle}\over
{ \langle v^\prime | v(A) \rangle}}\eqno{(8.10)}
$$
Gauge invariance is now a consequence of the gauge covariance of
the current operator ${\hat J}_\mu (x; A)$ and the cancelation
of the Wess-Zumino action between numerator and denominator. 
This formula says that the covariant current is the expectation
value of the charge current at the domain wall boundary. Although
${\hat J}_\mu (x; A)$ is local as an operator, the matrix element
between the two states $\langle v^\prime$ and $| v(A) \rangle$
induces the expected nonlocality in the covariant current functional
reflecting the integration of a massless fermionic degree of freedom. 

The formulae
would look slightly nicer had we rescaled ${\hat H} (A)$ by 2.
The standard definition of the overlap Dirac operator [\ovlapdirac] indeed has
the extra factor of 2 removed.

\vskip .5cm
\leftline{{\bf 9. Further improvement of the phase choice: two dimensions.}}
\vskip .5cm

At this point, to simplify the analysis, 
we restrict ourselves to two dimensions. At the conceptual level, 
nothing is lost by this restriction.

Consider, for fixed $y$, the quantity 
$$\epsilon_{\mu\rho} \nabla_\nu^{y-} S_{\rho\nu} (x,y;A)
\eqno{(9.1)}$$
This quantity can be viewed as an abelian noncompact vector potential
on the hypercubic lattice with varying argument $x$, 
defined so that now the backwards derivatives
induce gauge transformations.  
Equation (7.12) says that the abelian field strength associated
with this vector potential vanishes everywhere. Therefore,
the vector potential is pure gauge, which means there exists
a gauge function $\chi (x,y;A)$ such that
$$
\epsilon_{\mu\rho}\nabla_\nu^{y-} S_{\rho\nu} (x,y;A)=
\nabla_\mu^{x-} \chi (x,y;A)\eqno{(9.2)}$$
$\chi$ can be calculated starting from the site $x=y$, where $\chi$
is set to zero and getting to any other point
by summing  terms
$\epsilon_{\mu\rho}\nabla_\nu^{y-} S_{\rho\nu} (x,y;A)$ along
a chosen path. The vanishing of the above ``abelian field strength''
tells us that the result is independent of which particular path
we chose. $\chi$ is fixed uniquely up to a constant in $x$. This
constant can still depend on $y$ and $A$. The free constant
can be fixed by requiring $\chi$ to vanish at any fixed $y$ and
$A$ when $x$ is taken to $\infty$. With this, $\chi$ is uniquely
determined. 

Explicitly, we construct $\chi$ as follows: We choose
some path of minimal length in the lattice ``Manhattan'' metric. 
At large distances from $y$ the so constructed $\chi(x,y;A)$ will
have a vanishing gradient in $x$, so will become constant in $x$,
but possibly $y$ and $A$ dependent. This constant is approached
exponentially as $x$ increases. 
We redefine $\chi(x,y;A)$ by subtracting this constant (different
for each $y$ and $A$) from the $\chi(x,y;A)$ we have. The above
equation is still obeyed but now we see that  $\chi(x,y;A)$ is bilocal.
The entire construction only involved quantities invariant under
$A\to A^{(\alpha)}$, so $\chi(x,y;A)$ is also gauge invariant.
We now rewrite the last equation in the form:
$$
\nabla_\nu^{y-} S_{\mu\nu} (x,y;A) = \epsilon_{\mu\sigma} \nabla_\sigma^{x-}
\chi (x,y;A)\eqno{(9.3)}$$
The above construction is unique
and therefore must 
maintain the discrete symmetries of the original lattice
Schwinger term. So, we can write the new Wess-Zumino action in the form
$$
i\Phi^{\rm new} (\alpha ,A) =-
{1\over 2} \sum_{x,y} F_{12} (x) [ \int_0^1 dt
\chi (x,y;tA) ] \alpha (y) \eqno{(9.4)}
$$
This realizes the expectation to make it explicit that the coefficient
of $\alpha$ in the Wess-Zumino action is gauge invariant.

From equation (9.3) we expect $\chi(x,y;A)$ to be a total $y$-divergence
of some other local functional of $A$. 
If this were literally true the finite $y$-difference operation
could be thrown over to act on the $\alpha(y)$ factor in (9.4).
But then it would be evident that there exists a functional $\phi (A)$
such that $\Phi^{\rm new} (\alpha ,A)=\phi (A^{(\alpha)}) - \phi (A)$ 
and an additional phase redefinition by $\phi(A)$ would restore
gauge invariance. So, there must be an obstruction to writing
$\chi(x,y;A)$ as a total $y$-divergence. It is only this obstruction
that stands between our present phase choice and full restoration
of gauge invariance.

Let us focus therefore on the $y$-dependence of $\chi$.
What can stop $\chi (x,y ;A)$ from being a total
$y$-divergence of another local object is
that $\sum_y \chi(x,y;A) \ne 0$. A priori this is possible; there
is nothing to prohibit, for example, 
$\chi(x,y;A)={\rm Const.}\times \delta_{xy}$.
On the other hand, if the sum over $y$ of $\chi (x,y;A)$ were identically
zero, there would be many ways to write it as a total $y$ divergence. A
subset of these ways produces a bilocal current; this is the representation
we are after. 

So, we consider the quantity
$$
\sum_y \chi (x,y;A)=b(x;A)\eqno{(9.5)}$$
Because of equation (9.2) $b$ 
is $x$-independent. On the other hand,
$b(x,A)$ is also a local functional of $A$. Therefore, $b(x,A)$ must
be just a constant number:
$$
b(x;A)=b\eqno{(9.6)}$$
Hence, the obstruction has boiled down to the value of one
constant. 
The constant number $b$ came from the uniquely defined $\chi$,
which in turn came from the Schwinger term. Since $b$ is
independent of $A$, we can determine it at $A=0$. There we have
full translational invariance so that 
$S_{\mu\nu} (x,y;0) = S_{\mu\nu} (x-y)$ and similarly for $\chi$.
We go to Fourier space, and denote the Fourier transforms by
tildes. 
The antisymmetry of the Schwinger term implies that 
${\tilde S}_{\mu\nu} (0) = \epsilon_{\mu\nu} s_0$. Now,  we easily
deduce that $s_0 =b$. Hence:
$$
b= {1\over 2} \epsilon_{\mu\nu} \sum_x S_{\mu\nu} (x,y;0)\eqno{(9.7)}
$$

Let us now define a bilocal quantity $\psi(x,y;A)$ with $b$ subtracted:
$$
\psi (x,y; A)= \chi (x,y ;A) - b\delta_{x,y}\eqno{(9.8)}$$
Now, by construction, we have 
$$
\sum_y \psi (x,y;A)\equiv 0\eqno{(9.9)}$$
Also, from the properties of $\chi$ we know that the sum over $y$, restricted
to a large square surrounding $x$ fixed in its interior, 
and far from all edges of the square, converges exponentially to zero
as the box expands further. 
We want to convince ourselves that
his implies that there exists a bilocal functional with one index,
$\chi_\rho (x,y;A) $ such that
$$
\psi (x,y;A) =\nabla_\rho^{y-} \chi_\rho (x,y;A)\eqno{(9.10)}$$
First, fix $x$, and focus on the $y$ dependence only. Think
about $\psi(x,y;A)$ as an abelian noncompact two dimensional
field strength which is nonzero in some vicinity of a fixed
point ($x$). The sum condition on $y$ means that there is zero
total flux through the system. We simply want to find
the vector potential producing this field strength. To find
this vector potential we set up a maximal axial gauge tree
with origin at $y=0$. The axial gauge tree
is depicted in figure 1. In this gauge it determines 
the vector potential from
the field strength. The sums one needs
to do converge as a result of the locality
of $\psi (x,y;A)$. The question is whether
the resulting vector potential is still bilocal in $x$ and $y$.
For this we need the condition $\sum_y\psi (x,y;A)\equiv 0$. 
Next, for fixed $x$, 
we average over all possible trees with the above structure
to restore symmetries. Up to a factor of $\epsilon_{\sigma\rho}$
we get our $\chi_\rho (x,y;A)$ fields. 

\figure{1}{\captionone}{3.5}

Therefore,
$$
\chi (x,y;A) =\nabla_\nu^{y-} \chi_\nu (x,y;A) + b\delta_{xy}
\eqno{(9.11)}$$
With this, the Wess-Zumino action can be rewritten as:
$$
i\Phi^{\rm new} (\alpha ,A) =
{1\over 2} \sum_{x,y} F_{12} (x) [ \int_0^1 dt
\chi_\nu (x,y;tA) ] \nabla_\nu^{y+} \alpha (y) -{b\over 2} 
\sum_x F_{12} (x) \alpha (x)\eqno{(9.12)}$$
The first term above can be rewritten as
$$\eqalign{&
{1\over 2} \sum_{x,y} F_{12} (x) [ \int_0^1 dt
\chi_\nu (x,y;tA) ] \nabla_\nu^{y+} \alpha (y)=\cr
&{1\over 2} \sum_{x,y} F_{12} (x) [ \int_0^1 dt
\chi_\nu (x,y;tA) ] [A^{(\alpha)}_\nu (y)-A_\nu (y)]}\eqno{(9.13)}$$

By construction, $\chi_\nu (x,y;A)$ is invariant
under $A\to A^{(\alpha )}$. We therefore redefine our phase again:
$$
|v^{\rm new} (A) \rangle \to |v^{\rm final} (A) \rangle = 
e^{{1 \over 2}\sum_{x,y} F_{12} (x)  A_\nu (y) [ \int_0^1 dt
\chi_\nu (x,y;tA) ]} |v^{\rm new} (A) \rangle\eqno{(9.14)}$$
The new phase adjustment, just as the previous one, was by
a local quantity. The final form of the Wess-Zumino lattice functional
has been fine tuned now to its simplest possible structure,
exhibiting only the $b$ constant, the single irremovable
obstruction to full gauge invariance. 
$$
i\Phi^{\rm final} (\alpha ,A) = -{b\over 2} 
\sum_x F_{12} (x) \alpha (x)\eqno{(9.15)}$$
For each right handed fermion of charge $e_{Ra}$ one gets a factor
$+e_{Ra}^2$ and for each left handed fermion of charge $e_{Lb}$
one gets a factor $-e_{Lb}^2$. One charge factor
is associated with $F$ and another with $\alpha$. The constant
$b$ can be computed in the free theory. The contributions
of all fermions add up and perturbative anomalies cancel if
the total $b$ vanishes. 
If perturbative anomalies cancel, the Wess-Zumino action vanishes and
gauge invariance is fully restored, as there were no other
sources of gauge violation. 

On the other hand, if anomalies do not cancel, it is impossible to
find an additional phase redefinition, by some local functional
$\phi (A)$ such that the gauge dependence be made to disappear
by $\Phi^{\rm final} (\alpha , A) = \phi (A^{(\alpha )})-\phi (A)$.
One can address this directly on the lattice, but it is easier
to just observe that if it were possible, one could take a naive
continuum limit and eliminate the anomaly also in the continuum.
This is well known to be impossible. 

\vskip .5cm
\leftline{{\bf 10. Higher dimensions}}
\vskip .5cm

Essentially the same story goes through in higher (even) dimensions
than two. The considerations used in section 9  for constructing
the fields $\chi$ and $\chi_\nu$ were those of ordinary abelian
noncompact lattice field theory only accidentally. 
Going to higher dimensions involves more indices and 
considerations in auxiliary noncompact lattice abelian gauge theory,
generalized to local sub-hypercubes of arbitrary dimension:
From sites, links plaquettes one needs to go to three-cubes,
four-cubes and so forth.  As is well
known, this generalization involves antisymmetric tensors of higher rank,
the rank being given by the dimension of the sub-hypercube.
All variables are non-compact, real numbers. Gauge transformations
and the relation between variables and gauge invariant ``field strengths''
are all linear and involve lattice finite difference operations. 
The basic field variables are defined on $p$ dimensional
sub-hypercubes, and have a gauge invariance under variables
that live on their boundary, i.e. on $p-1$ dimensional sub-hypercubes.
The gauge invariant variables that would enter an action
are defined on $p+1$ dimensional sub-hypercubes and are 
sums with orientation depending signs of the basic variables on the
respective boundaries. Ordinary gauge theory has $p=1$, where
the variables live on links, the gauge transformation functions on
sites ($p=0$) and the field strength variables on plaquettes
($p=2$). The two basic questions one needs to deal with are when is
a variable pure gauge and how to recover a variable from a known
field strength. An entire hierarchy labeled by $p$ is
employed. The highest $p$ is determined by the dimension $d$. 
The increase in tedium can be substantial, but an attempt to formalize it  
to all dimensions is described in [\fujiwara]. The entire
auxiliary geometric structure is hidden by developing a finite difference 
calculus on antisymmetric lattice tensor fields. 

Combining the existence of the current $j_\mu (x;A)$
established in section 7 
with the symmetry properties of the 
adiabatic phase choice proven in section 13
and with the result of [\fujiwara] on local abelian lattice
BRS cohomology, 
we learn that most of the gauge dependence of $j_\mu (x; A)$ can 
be eliminated by the following decomposition:
$$\eqalign{
j_\mu (x &; A)= k_\mu (x;A) +\cr
& a_0 \overline{ \epsilon_{\mu , \nu_1 ,....\mu_k , \nu_k}
 A_{\nu_1 }(x) F_{\mu_2 \nu_2 }(x+{\hat \nu}_1 )
...F_{\mu_{d_2} \nu_{d_2} }(x+{\hat \nu}_1+{\hat \mu}_2+{\hat \nu}_2 +...
+{\hat \mu_{d_2-1}}+{\hat \nu}_{d_2-1}) }\cr + &j^{g.i.}_\mu (x; A)}
\eqno{(10.1)}$$
Any $\mu,\nu$ index repeated more than once is summed over in (10.1).
The top bar
indicates that the quantity needs to be averaged with respect to
lattice symmetries to comply with those of the current 
$j_\mu (x; A)$. The current $j^{g.i.}_\mu (x; A)$ is both gauge
invariant and local. The current $k_\mu (x;A)$ is not gauge invariant
but has zero divergence. 
The $a_0$ term is the single term that is simultaneously not gauge
invariant and not divergence-less. 
The entire divergence of the current $j_\mu (x ; A)$ 
is contained in the $a_0$ term together with 
the $j^{g.i.}_\mu (x; A)$ term. 
The constant $a_0$ can be determined in perturbation theory,
and is proportional to the coefficient of the perturbative
anomaly. It again comes from the Schwinger term, but this time
one needs to expand to order $d_2-1$ in the field strength.

If we now redefine the phases as above, we see that we are left
with a Wess-Zumino functional $\Phi (\alpha ,A)$ of the following
form:
$$\eqalign{
&\Phi (\alpha, A) =i a_0 \sum_x \alpha(x) \cr
&\overline{ \epsilon_{\mu_1 , \nu_1 ,....\mu_k , \nu_k}
 F_{\mu_1 \nu_1 }(x) F_{\mu_2 \nu_2 }(x+{\hat \mu}_1 +{\hat \nu}_1 )
...F_{\mu_{d_2} \nu_{d_2} }(x+{\hat \mu}_1 +
{\hat \nu}_1 +...
+{\hat \mu}_{d_2-1}+{\hat \nu}_{d_2-1}) }}
\eqno{(10.2)}
$$
When there are several fermions of charges $q$ and we choose for each
a phase definition that produces this minimal form, we shall
restore gauge invariance if the sum of all the 
$a_0 \times q^{d_2 +1}$ constants
vanishes. This is the standard anomaly cancelation condition. 
When $d_2$ is even we can discuss only one handedness and cancelations
occur between positive and negative charges. When
$d_2$ is odd, fermions of both handednesses 
must participate to get a cancelation. 

\vskip .5cm
\leftline{{\bf 11. 
Lattice effective actions for anomalous theories in four dimensions}}
\vskip .5cm
An anomalous chiral abelian gauge theory with an ultraviolet cutoff $\Lambda$
can be viewed as an effective theory describing continuum physics
at momenta low relative to $\Lambda$. Assuming a single Weyl fermion,
this effective theory has predictive power only if the 
physical photon mass, $m^{\rm ph}_\gamma$, obeys:
$$1>> {{m^{\rm ph}_\gamma}\over
\Lambda}  \ge {{(e^{\rm ph})^3}\over {64 \pi^3 }} \eqno{(11.1)}
$$
This bound is derived by perturbative power counting arguments in
the continuum without specifying the cutoff [\preskill]. 

The work in the present paper presents a way to check (11.1) 
outside perturbation theory, with a lattice cutoff. We take
the overlap with one of the phase definitions presented earlier
and including it into the pure gauge action defines the path integral
from which the photon mass in lattice units can be extracted.
We should find that we cannot make the photon mass smaller than
the bound, no matter what we do (within reason)
with the bare photon mass parameter, $m_\gamma$.  

It would be interesting to know if the bound depends on whether
we use the simple adiabatic phase choice, or any of the subsequent
improvements. The simplest expectation is that the sensitivity
of the bound on whether we pick the simplest adiabatic
phase choice or any of the improved ones be low: The phase choices
differ only by local terms added to the gauge action. For all
phase choices the Wess-Zumino functional is linear in the gauge 
transform field $\alpha$
and one can use in perturbation theory Stueckelberg's trick
of making the gauge field $\alpha$ dynamical, thus endowing
the fermion sector with a new gauge invariance, where $\alpha$
transforms by shifts under the new gauge transformation. This trick
is employed in the perturbative analysis of Preskill and Wise. 

Unfortunately, a practical numerical simulation would have to
deal with the complex measure in the path integral. There
are no known generic ways to deal with extensive phases, so
one would need to invent a specific procedure for this case.
Another technical obstacle is the numerical local bound
on the abelian field strength, which might force the simulations
to impractically large lattices. 

If anomalies do cancel, one should find that it is easy to make the 
photon massless as $m_\gamma$ is taken to zero. This seems quite
obvious in the case one adopts the perfect phase choice. It would
be very interesting to see what happens if one adopts
the less perfect adiabatic phase choice we started with. 
This question, posed here in the noncompact context, is somewhat
similar to tests of ``gauge averaging'' in compact formulations. 
A more precise formulation of gauge averaging in the noncompact
framework seems difficult: 
Gauge restoration in the infrared by
gauge averaging is a mechanism (due to [\fnn]) 
that works in a strong coupling expansion. In this strong coupling
expansion 
the compactness of the gauge group plays a central role.

Equation (11.1) can be read in two 
additional ways: as an upper bound on the ultraviolet cutoff
and as an upper bound on the physical coupling constant. 
The difference between the two effective theories, one with
canceled anomalies and the other with uncanceled anomalies
can be phrased as follows: Let $m^{\rm ph}$ denote a typical 
low physical scale where the effective Lagrangian applies.
We assume ${{m^{\rm ph}}\over \Lambda} << 1$. 
The physical coupling is bounded from above in either case.
The bound depends on the ultraviolet cutoff, just like in
the Higgs mass bound case [\higgsbound]. If anomalies do not cancel we
have a bound of the following type:
$$
e^{\rm ph} \le c_1 \left ( 
{{m^{\rm ph}}\over \Lambda } \right )^{1\over 3}\eqno{(11.2)}$$
If anomalies do cancel the limitation is far less severe:
$$
e^{\rm ph} \le {{c_2}\over {\log 
\left ( {\Lambda \over {m^{\rm ph}}}\right )}}
\eqno{(11.3)}$$ 
Equations (11.2) and (11.3) probably are the most physical way
to express the difference between an anomalous and a 
non-anomalous abelian chiral gauge theory in four space-time 
dimensions. 

This paper can easily be generalized to include an abelian
group which has several $U(1)$ factors. The various individual
and mixed anomalies would work out just as in the continuum.
Let us demand that the``triviality'' bound on the couplings be 
logarithmic and take the four $U(1)$ factors
one would get from the gauge group of the minimal standard model,
gauge them, and introduce the fermion content of one generation.
We shall find, as in the continuum, that the ratios between all
hypercharges are fixed to their
known values by anomaly cancelation conditions.

\vskip .5cm
\leftline{{\bf 12. Single particle version of the adiabatic phase choice}}
\vskip .5cm

Since 
$$
{\hat H}(A) = {\hat a}^\dagger \epsilon(A) {\hat a} + N_v 
\eqno{(12.1)}$$
the overlap is determined by the matrix 
\footnote{${}^{f_3}$}{This matrix is closely related
to the overlap Dirac 
operator $D_o ={1\over 2} (1+\gamma_{d+1} 
\epsilon (A))$ [\ovlapdirac].}
$\epsilon (A)$. The
eigenstates of this matrix are ${\vec v}_i (A)$ and
${\vec w}_i (A)$ for eigenvalues
$\mp 1$ respectively. Note that the ${\vec v}_i (A)$
are not necessarily eigenstates of $H_W(A)$, but they exactly span the
negative energy subspace of $H_W(A)$.

The ground state of ${\hat H}(A)$ has all the states
${\vec v}_i (A)$ occupied and the rest empty. Thus the
information contained in the ground state $|v(A)\rangle$
is the same as that contained in the rectangular matrix
$v=({\vec v}_1 (A),{\vec v}_2 (A),....,{\vec v}_{N_v} (A) )$,
where $N_v$ is the total number of negative energy states. 
Similarly we define the matrix $w$ made out of all positive
energy states. Also, for the reference system, we introduce
the same quantities, all with a prime superscript. 

The overlap corresponding to right handed chiral fermions is
$$
\langle v^\prime | v(A)\rangle = \det M_R ,~~~~ M_R = v^{\prime\dagger} v(A)
\eqno{(12.2)}$$
and that to left handed chiral fermions is
$$
\langle w^\prime | w(A)\rangle = \det M_L, ~~~~ M_L = w^{\prime\dagger} w(A)
\eqno{(12.3)}$$

The matrix $v(A)$ is not fully defined because
we can unitarily mix the negative energy states. 
If we make some arbitrary choice, the corresponding state
$|v(A) \rangle$ will not have an adiabatically defined phase.
Note that most of the details of a unitary mix 
are unimportant, only the collective effect on the phase of
the second quantized state $|v(A) \rangle$ matters. 

Let us start
from some initial choice of single particle states ${\tilde v} (A)$
and ${\tilde w}(A)$. We are restricting our attention to the set of $A$'s for
which
$H^2 (A)$ is bounded from below by some small positive number. 
The lower bound (2.13)
on $H^2 (A)$ means that no state can cross between the negative energy
and positive energy groups of states. 
The states $v(A)$ and $w(A)$ have adiabatic phases and are given by
$$
v(A) = {\tilde v}(A) {\cal O}^\dagger_v (A),~~~~
w(A) = {\tilde w}(A) {\cal O}^\dagger_w (A) \eqno{(12.4)}$$

The overlaps with the two phase choices are related by:
$$
\langle v^\prime | v(A) \rangle = \det v^{\prime\dagger} v(A)=
\det {\tilde v}^{\prime \dagger} {\tilde v}(A) \ \det {\cal O}_v^\dagger (A),
\eqno{(12.5)}$$
where, by convention $v^\prime = {\tilde v}^\prime$.
There always are equal numbers of negative
energy and positive energy states.
The matrices ${\cal O}$ are defined to be unitary and 
required to be equal to unit matrices for
$A=0$. They are completely fixed by the differential
equation below:
$$
{{d {\cal O}_v (tA) }\over {dt}} = 
{\cal O}_v (tA) {\tilde v}^\dagger (tA) {{d \tilde v (tA)}\over {dt}},~~ 
{{d {\cal O}_w (tA) }\over {dt}} = 
{\cal O}_w (tA) {\tilde w}^\dagger (tA) {{d \tilde w (tA)}\over {dt}},~~ 
\eqno{(12.6)}$$
In turn, this implies,
$$
v^\dagger (tA) {{d v (tA )}\over {dt}} =0,~~~~
w^\dagger (tA) {{d w (tA )}\over {dt}} =0
\eqno{(12.7)}$$
which, in particular, makes $|v(A)\rangle$ have adiabatic phases:
$$
\langle v(tA )| {{d v(tA)}\over {dt}}\rangle = 
\det \left [ v^\dagger (tA) {{d v (tA )}\over {dt}}\right ]
\eqno{(12.8)}$$

Our objective is to relate the adiabatic phase choice to the BW one,
defined by
$$
\langle {\tilde v}(0) | {\tilde v}(A) \rangle >0,~
\langle {\tilde w}(0) | {\tilde w}(A) \rangle >0,~~~|v(0)\rangle = 
{\tilde v} (0) \rangle,~|w(0)\rangle = 
{\tilde w} (0) \rangle \eqno{(12.9)}$$
When found, this 
relation will allow us to extend to the adiabatic phase choice 
some symmetry
properties proven before [\ovlapb] for the BW phase choice.

The procedure to find the relation between the two phases
relies on the fact that in both cases one uses the $A=0$
as a reference point. The reference points
for the two phase choices can be made the same. 
The adiabatic evolution takes
one from reference states, unitarily, to the adiabatic 
states at arbitrary $A$. The unitary matrix, $K(A)$ [\dubna], 
which does this was introduced by
Kato [\kato] and is defined below:
$$
K(A)= v(A){\tilde v}^\dagger (0) + w(A){\tilde w}^\dagger (0)=
{\tilde v}(A){\cal O}_v^\dagger (A) {\tilde v}^\dagger (0) + 
{\tilde w}(A){\cal O}_w^\dagger (A) {\tilde w}^\dagger (0)
\eqno{(12.10)}$$
By definition, $K(0)=1$. The rest of $K$ is defined
by deriving a first order evolution equation.
For it one employs the projectors
$$
P(A)= v(A) v^\dagger (A) = {{1- \epsilon(A)}\over 2},
~~~~1-P(A)=w(A) w^\dagger (A)={{1+ \epsilon(A)}\over 2}
\eqno{(12.11)}
$$
The projectors are defined unambiguously, by the subspaces
they project on. Kato observed that 
$$
{{dK(tA)}\over{dt}}= [{{dP(tA)}\over{dt}} , P(tA) ] K(tA)
\eqno{(12.12)}$$
This formula is easy to prove.

By equation (12.4) the relation between the BW phase choice and
the adiabatic phase choice is contained in the matrix 
${\cal O}_v^\dagger (A)$. But there is no independent
definition of the matrix ${\cal O}_v^\dagger (A)$.
To trade the matrix ${\cal O}_v^\dagger (A)$
for the matrix $K(A)$ which is independently defined 
we write:
$$
{\tilde v}^\dagger (0) {\tilde v}(A) {\cal O}_v^\dagger (A) =
{\tilde v}^\dagger (0) K(A) {\tilde v}(0)\eqno{(12.13)}$$
Taking determinants we get:
$$\eqalign{
\det &{\cal O}_v^\dagger (A) \ \det [ {\tilde v}^\dagger (0) {\tilde v}(A) ]
= \det \left [ 1-{\tilde v}^\dagger (0) [1-K(A)]{\tilde v}(0) \right ]=
\cr & \exp \left \{ - \sum_{m=1}^\infty {1\over m} 
Tr \left [ {\tilde v}^\dagger (0) [1-K(A)]{\tilde v}(0) \right ]^m \right \}
=
\exp \left \{ - {\sum_{m=1}^\infty {1\over m}
Tr \left [  [1-K(A)] P(0) \right ]^m} \right \} }
\eqno{(12.14)}$$
So, we learn that:
$$
\det {\cal O}_v^\dagger (A) \ \langle {\tilde v}(0)| {\tilde v} (A) \rangle =
\det \left [ 1-P(0) + K(A) P(0) \right ]=
\det \left [ 1-P(0) + P(0) K(A) \right ]
\eqno{(12.15)}$$

Since, by definition of the BW phase choice, the factor
$\langle {\tilde v}(0)| {\tilde v} (A) \rangle$ is positive,
the determinants after the first and second equality have
the same phase as $\det {\cal O}_v^\dagger (A)$. Thus we arrive at
a formula relating the overlap with BW phase choice to the overlap
with an adiabatic phase choice:
$$
\langle v^\prime | v(A)\rangle =
\langle {\tilde v}^\prime | {\tilde v}(A) \rangle 
{{\det \left [ 1-P(0) + K(A) P(0) \right ]}
\over {|\det \left [ 1-P(0) + K(A) P(0) \right ]|}}
\eqno{(12.16)}$$
By definition, $\langle v^\prime | = \langle {\tilde v}^\prime |$. 
So long the matrix  $1-P(0) + K(A) P(0)$ is not singular the two
phase choices are smoothly and locally related. 

In numerical work one needs more explicit expressions for the
various matrix elements in Fock space we encountered. For completeness
I include below an overlap ``master formula'', provable by
purely combinatorial means [\ovlapb, \dubna]. 
This formula tells how to transcribe
all Fock space expressions one may need into single particle
language. 

Assume that we have the rectangular matrix $v$ as above and another
matrix of similar structure $u$. In applications $u$ can be taken
as $v^\prime$ or as $v$. The main assumption is that the columns of
$u$ are linearly independent and their number is the same as in 
$v$. Let $P_v , P_u$ denote the projectors on the
subspaces spanned by the columns:
$$
P_v = vv^\dagger,~~~P_u = uu^\dagger,~~~v^\dagger v =1,~~~
u^\dagger u=1\eqno{(12.17)}$$
Let ${\hat a}^\dagger X {\hat a} = {\hat X}$ be a bilinear operator.
Then:
$$\eqalign{
&\langle u| v\rangle = \det (u^\dagger v)\cr
\langle u |{\hat X} |v \rangle =& -\det (u^\dagger v)
Tr \left [X P_v {1\over {1-P_u -P_v }} \right ] =-\det (u^\dagger v)
Tr \left [P_u X {1\over {1-P_u -P_v }} \right ]}\eqno{(12.18)}$$

For $u=v$ we can use $P_v (1-2P_v) = -P_v$ to get
$$
\langle v |{\hat X} | v\rangle = Tr (XP_v )\eqno {(12.19)}$$
The expression ${1\over {1-P_u -P_v }}$ comes accompanied by
$\det (u^\dagger v )$ and reflects the nonlocality in a theory
with massless fermions when $u=v^\prime$. 
However, for $u=v$
the nonlocality disappears. This is how the Wess-Zumino action
becomes local in the first place.
Equation (12.18) 
with $u=v^\prime$ makes it explicit that all gauge breaking comes 
from the fermion determinant factor $\langle u | v\rangle$. 
Setting $u=v$ typically produces expressions with naive gauge
transformation properties.  

Another combinatorial fact, proven in [\ovlapb] is that even
for $u\ne v$ Wick's theorem still holds. By this I mean that
the $u-v$ matrix element of any string of fermionic creation
and annihilation operators is given by sums of products
of $\langle u | a^\dagger_\mu (x) a_\nu (y) | v\rangle$,
where each matrix element is divided by the overlap $\langle u |
v\rangle$. Therefore, equation (12.18) provides enough information
to evaluate $\langle u |{\hat X}|v\rangle$ for any ${\hat X}$,
not just bilinear ones.

\vskip .5cm
\leftline{{\bf 13. Symmetry properties of adiabatic phase choice}}
\vskip .5cm

Relation (12.16) is now used to show that left and right handed
chiral fermions of the same charge have determinants of opposite
phase. First relation (12.16) has to be generalized to
the opposite handedness. This is easy: replace all $v$'s by $w$'s,
which induces also replacing the projector $P(A)$ by $1-P(A)$.
The matrix $K(A)$ remains unchanged, because the defining first
order differential equation and the initial condition have not
changed. Also,
$\det K(A) =1$ because it is so at $A=0$ and the evolution does
not change the determinant. Taking a complex conjugate of (12.16)
we get:
$$\eqalign{
\langle v^\prime | v(A)\rangle^*
=&\langle {\tilde v}^\prime | {\tilde v}(A) \rangle^*
{{\det \left [  1-P(0) + P(0) K^\dagger (A) \right ]}
\over
{|\det \left [  1-P(0) + P(0) K^\dagger (A) \right ] |}}=\cr
&\langle {\tilde v}^\prime | {\tilde v}(A) \rangle^*
{{\det \left [ [ 1-P(0)]K(A) + P(0) \right ]}\over
{|\det \left [ [ 1-P(0)]K(A) + P(0) \right ]|}}=\cr
&
\langle {\tilde v}^\prime | {\tilde v}(A) \rangle^*
{{\det \left [P(0)+K(A) [ 1-P(0)] \right ]}\over
{|\det \left [P(0)+K(A) [ 1-P(0)] \right ]|}}
}\eqno{(13.1)}$$

We know from [\ovlapb] that
$$
\langle {\tilde v}^\prime | {\tilde v}(A)\rangle^* =
\langle {\tilde w}^\prime | {\tilde w}(A)\rangle\eqno{(13.2)}$$
With this substitution we get
$$
\langle v^\prime | v(A) \rangle^* = \langle w^\prime | w(A) \rangle
\eqno{(13.3)}$$
Therefore, one of the key properties of the BW phase choice,
namely that the imaginary part of the induced action switches
sign when handedness is switched also holds with the adiabatic
phase choice. Also, the real parts of the induced action 
are identical for the left and right handedness.

The above property is true in any dimensions. In some special
dimensions, like four, one can deal with fermions of only
one handedness and use conjugate representations for the
other handedness and thus avoid defining separately 
lattice fermions of left and right handedness. But, his would not work in
two dimensions for example. 

It is now not hard to see how all other symmetry properties of
the overlap with a BW phase choice hold also with the adiabatic
phase choice. One always uses the relation (12.16) and symmetry
properties of $H(A)$. The symmetry properties
of the matrix $H(A)$ are directly inherited by $K(A)$,
so long the free fermion ground state is invariant, making $P(0)$
invariant.

\vskip .5cm
\leftline{{\bf 14. Nonabelian case}}
\vskip .5cm
It is quite clear from the above that the abelian nature of the gauge
group enters in many places. The most technical
component in the above construction is the decomposition of
the current into a divergenceless piece, a topological piece and a
gauge invariant piece. For our abelian case 
this technical part could probably be done
more efficiently in Fourier space. But, when one contemplates a
generalization to the nonabelian case, coordinate space
might be better. In coordinate space the abelian current decomposition
formula comes from a study of local abelian lattice BRS cohomology. The
continuum BRS cohomology structure of nonabelian gauge theories
has been transcribed to the lattice in references [\latticebrs].
There one also dealt with cohomology issues, but not the local
version, which is relevant here. It is conceivable that on these
lines some generalizations to the nonabelian case can be found.
Even if this is done, one still has more work to do if one
wants to treat nonabelian lattice chiral gauge theories on lines
that closely generalize the abelian treatment of this paper.

\vskip .5cm
\leftline{{\bf 15. Summary}}
\vskip .5cm

In the particular case of noncompact chiral abelian gauge theories
at infinite volume it seems possible to formulate a lattice version
which for all purposes of principle has all the simplicity of 
a continuum formulation, only now holding nonperturbatively.
It is possible to make what was termed in the last section of [\ovlapb] 
a ``perfect'' phase choice
in this abelian noncompact case: restoration of gauge invariance is possible
if and only if perturbative anomalies cancel. In the anomalous
case one has an effective theory that looks very much
what one would write down based on perturbative arguments.
One could make concrete the meaning of the nonrenormalizability
bounds introduced by Preskill and Wise in the abelian case, this
time outside perturbation theory. Generalization of the perfect
phase choice to the nonabelian case seems difficult. On the other
hand, the construction in the abelian case is quite physical and
the concept of adiabatic evolution does generalize. 

Technically,
the physical aspects of the construction are much more transparent
in Fock space (second quantized language). This is not unexpected
in the context of field theory. But, the field theory is quadratic,
because we only quantize some auxiliary fermions in a fixed gauge
background. Therefore, all expressions can be easily transcribed
into single particle language. This language is needed in making
the expressions useful for numerical work. The translation
between the two languages is just a matter of some combinatorics. 

This paper does not shed new light on the question whether a perfect
phase choice should be viewed as fine tuning or not. Since we work
in the noncompact case the issue of gauge averaging a slightly
imperfect phase choice cannot be directly addressed. An indirect
approach was sketched, but seems difficult to implement computationally
at present.

If we think about the standard model, we now can put on the
lattice a theory with the fermion content of one or more
generations, in which we gauged the largest continuous 
abelian subgroup of $U(1)\times SU(2) \times SU(3)$ and
maintained exact (noncompact) gauge invariance. 
The entire intricate mechanism of anomaly cancelation
comes then into play. For example, the well known 
restrictions on physical charges apply now in a nonperturbative setting. 
So, we probably are one step closer to putting the entire minimal standard
model on the lattice.

\vskip .5cm
\leftline{{\bf Acknowledgments}}
\vskip .5cm
My research at Rutgers is partially supported by the DOE under grant
\# DE-FG05-96ER40559. 
\vskip .5cm
\leftline{{\bf References}}
\vskip .5cm
\item{\bf \ovlapa} R. Narayanan, H. Neuberger,
\Journal{\PLB}{302}{62}{1993};
\Journal{\PRL}{71}{3251}{1993};
\Journal{\NPB}{412}{574}{1994}.
\item{\bf \luscher} M. L{\" u}scher, hep-lat/9909150.
\item{\bf \ovlapb} R. Narayanan, H. Neuberger,
\Journal{\NPB}{443}{305}{1995}.
\item{\bf \dubna} H. Neuberger, hep-lat/9912013.
\item{\bf \kaplan} D. B. Kaplan, \Journal{\PLB}{288}{342}{1992}.
\item{\bf \fands} S. A. Frolov, A. A. Slavnov,
\Journal{\PLB}{309}{344}{1993}.
\item{\bf \candh} C. Callan, J. Harvey, \Journal{\NPB}{250}{427}{1984}.
\item{\bf \thooft} G. 't Hooft, \Journal{\PLB}{349}{491}{1995}.
\item{\bf \zinn} J. Zinn-Justin, {\it Quantum Field Theory and Critical
Phenomena} (Oxford University Press, 1993).
\item{\bf \preskill} J. Preskill, \Journal{\ANPH}{210}{323}{1991}.
\item{\bf \boulder}H. Neuberger, \Journal{\NPROC}{73}{697}{1999}. 
\item{\bf \bsimon} B. Simon, \Journal{\PRL}{51}{2167}{1983}.
\item{\bf \geom} H. Neuberger,
\Journal{\PRD}{59}{085006}{1999}.
\item{\bf \berry} M. Berry, \Journal{\PRL}{392}{45}{1984}.
\item{\bf \mybounds} H. Neuberger, hep-lat/9911004.
\item{\bf \gw} P. Ginsparg, K. Wilson,
\Journal{\PRD}{25}{2649}{1982}.
\item{\bf \daemi} S. Randjbar-Daemi, J. Strathdee, 
\Journal{\NPB}{466}{335}{1996}.
\item{\bf \kato} T. Kato, \Journal{\JPSJ}{5}{435}{1950}.
\item{\bf \chiuone} R. Narayanan, H. Neuberger, \Journal{\NPB}
{477}{521}{1996}.
\item{\bf \fuji} K. Fujikawa, \Journal{\PRL}{42}{1195}{1979}.
\item{\bf \shamir} Y. Shamir, \Journal{PRD}{57}{132}{1998}.
\item{\bf \ovlapdirac} H. Neuberger, 
\Journal{\PLB}{417}{141}{1998}.
\item{\bf \fujiwara} T. Fujiwara et. al., hep-lat/9906015.
\item{\bf \fnn} D. F{\" o}rster, H. B. Nielsen, M. Ninomiya,
\Journal{\PLB}{94}{135}{1980}.
\item{\bf \higgsbound} U. M. Heller et. al., 
\Journal{\NPB}{405}{555}{1993}.
\item{\bf \latticebrs} H. Neuberger, 
\Journal{\PLB}{175}{69}{1986}; \Journal{\PLB}{183}{337}{1987}.

\vfill\eject\bye